# Change in the order of magnetic transition in HoRhGe as probed by magnetoresistance and magnetocaloric studies

Sachin B. Gupta,[1] K.G. Suresh,[1,a] and A.K. Nigam[2]

[1]Department of Physics, Indian Institute of Technology Bombay, Mumbai - 400076, India

[2]Tata Institute of Fundamental Research, Homi Bhabha Road, Mumbai - 400005, India

## Abstract

Magnetoresistance and magnetocaloric properties of polycrystalline HoRhGe have been studied. This compound orders antiferromagnetically with a Neel temperature ($T_N$) of 5.5 K and undergoes a first order metamagnetic transition at 2 K. It shows a large negative magnetoresistance of 25% at $T_N$, for a field of 50 kOe. However, at 2 K and in lower fields, the magnetoresistance is found to be positive with a magnitude of about 12%, which is attributed to the first order metamagnetic transition. The compound also shows large magnetocaloric effect near its Neel temperature. Field dependence of magnetic entropy change also reflects the change in the nature of magnetic transition as the field is increased. The value of magnetic entropy change ($-\Delta S_M$) is found to be 11.1 J/kg K for a field change of 50 kOe. Field dependence of magnetoresistance and magnetocaloric effect clearly shows the change in the order of the metamagnetic transition with increase in field.

___________________________________________________________________

Corresponding Author Email: suresh@phy.iitb.ac.in

# Introduction.

Observation of many intriguing properties and the application potential has made the rare earth intermetallic compounds very interesting candidates in the area of magnetic materials. Of particular interest is the compounds which exhibit phenomena such as metamagnetic transitions, spin glass behavior, giant magnetocaloric effect (GMCE), giant magnetoresistance (GMR), intermediate valence behavior, superconductivity etc [1-6]. Among these, magnetocaloric effect (MCE) and magneoresistance (MR) have direct bearing on applications. Rare earth based intermetallic compounds are found to be the most potential magnetic refrigerant materials because of their many advantages over their ceramic counterparts. As far as GMR is concerned, usually it is observed in metallic multilayer systems, as a result of a field induced transition between antiparallel and parallel couplings between the two magnetic layers separated by a nonmagnetic layer. Apart from such artificial multilayer systems, there are naturally occurring intermetallic compounds, which have a crystal structure in which the magnetic ions remain in well separated layers. The mechanism of MR in such materials is found to be different from the usual negative MR seen in many intermetallic compounds where the applied/internal field suppresses the spin disorder and spin flip contributions to the resistivity.

Among the intermetallic compounds, RTX series (T= 3d/4d element, X= p block element) has received the attention of a large number of researchers in the past. Many compounds of this series are being re-investigated with the aim of identifying their potentials for applications such as magnetic refrigeration. RRhGe family belonging to this series is known to show many interesting magnetic properties. In order to find out their magnetocaloric and magnetoresistance behavior, we have made a detailed study of RRhGe compounds with various rare earths. Among these compounds, HoRhGe is found to show many interesting properties like field-induced metamagnetic transition, large negative magnetoresistance and relatively large magnetocaloric effect, with anomalous field dependencies.

# Experimental details.

The polycrystalline sample was synthesized by arc melting the constituent elements of at least 99.9% purity in a water-cooled copper hearth, under argon atmosphere. The sample was annealed for a week at 800 ℃ for better homogeneity. The room temperature powder x-ray diffraction (XRD) pattern was taken on a X'PERT PRO diffractometer using Cu Kα (λ=1.54 Å) radiation. The magnetization (M) and heat capacity (C) measurements were performed using a Physical Property Measurement System (Quantum Design, PPMS-6500). The electrical resistivity (ρ) measurement with and without field was also performed on the PPMS using standard four probe technique, applying an excitation current of 150 mA.

## Results and discussion.

The Rietveld refinement of the XRD pattern collected at room temperature shows single phase nature of the compound. It is found to crystallize in the TiNiSi type orthorhombic structure with space group Pnma (No. 62). The lattice parameters are found to be a=6.8519(3) Å, b= 4.2711(2) Å, c= 7.5118(3) Å. These are in agreement with the values reported by Hovestreydt *et al.*[7].

The temperature dependence of dc magnetic susceptibility (DCS), recorded in a field of 500 Oe and in the temperature range of 1.8-300 K is shown in figure 2. The Neel temperature ($T_N$) estimated from dχ/dT versus T plot, is found to be 5.5 K. The magnetic ordering temperature observed from neutron diffraction study [8] was found to 4.6 K. The modified Curie-Weiss fit, $\chi^{-1} = (T - \theta_p)/C_m$, (see figure 2) yields the effective magnetic moment ($\mu_{eff.}$) of 11.08 $\mu_B$/Ho$^{3+}$ and the paramagnetic Curie temperature ($\theta_p$) of -1.9 K. It may be noted that the theoretical free ion moment of Ho$^{3+}$ is 10.6 $\mu_B$. The fact that the value of $\theta_p$ is less than $T_N$ suggests that the strength of the antiferromagnetic ordering is not very strong in this compound, as is also evidenced by other measurements described below.

Figure 3(a) shows the magnetization isotherms at some selected temperatures obtained for fields up to 70 kOe. At 2 K and for low fields, the magnetization changes linearly as a function of field, reflecting the antiferromagnetic nature of the compound, while on increasing field beyond a certain value (≈9.57 kOe), a sharp increase in magnetization occurs. This change in magnetization is due to the field-induced metamagnetic transition. The critical field ($H_C$) was evaluated from the peak position of dM/dH plot as shown in the inset of figure 3(a). At higher fields, a change in the curvature of the magnetic isotherms occurs, which suggests that the antiferromagnetic interaction is not strong. However, the compound shows only a weak saturation trend even in high fields.

To check the nature of the magnetic transition in this compound, the magnetization data were plotted in the form of Arrott plots [see figure 3(b)]. According to Banerjee criterion [9], the magnetic transition is second order if all the Arrott plots have positive slope, while it is of first order if some of the Arrott plots have negative slopes. One can see from figure 3(b) that for T< $T_N$, the Arrott plots have initial negative slope, indicating the first order nature of the transition. The first order transition in HoRhGe has also been confirmed by Landau coefficient estimated by the fit of the equation of state, $\mu_0 H = c_1(T)M + c_3(T)M^3 + c_5(T)M^5$ to the magnetization data [see inset of figure 3(b)]. The value of $c_3(T)$ is negative in low fields (≈ 10 kOe) and below $T_N$, and is positive for higher fields and for T≥$T_N$. It has been reported [10] that the order of magnetic transition is related to the sign of $c_3(T)$. It is first order if the sign of $c_3(T)$ is negative, whereas it will be second order for a positive $c_3(T)$. Hence, it is clear from the Arrott plots and the equation of state that this compound shows first order field induced metamagnetic transition in low fields and below $T_N$.

In order to understand the magnetic state of HoRhGe in more detail, the heat capacity measurement was performed with and without field in the temperature range of 2-100 K. The peak broadens on the application of the field and gets shifted to higher temperature (see inset of figure 2), revealing that HoRhGe is only weakly antiferromagnetic. Electrical resistivity and MR measurements were also performed on this compound. The magnetoresistance percentage (MR%) has been calculated from the field dependence of the resistivity given as

$$MR\% = \left(\frac{\rho(T,H) - \rho(T,0)}{\rho(T,0)}\right) \times 100. \qquad (1)$$

One can see from the temperature dependence of MR [see figure 4(a)] that the magnitude of MR increases with decrease in temperature and reaches the maximum at $T_N$. The maximum negative value obtained for a field change of 50 kOe is found to be -25 %. However, the most interesting observation is the positive MR seen at 2 K for low fields in the MR versus H plots [see figure 4(b)]. At 2 K, the positive MR initially increases with field, reaching its maximum value of about 12% at ≈9.5 kOe. For further increase in the fields, the MR value decreases and becomes almost zero at 50 kOe. Therefore, one can note that the magnitude of MR at 2 K increases with field upto fields equal to $H_C$, thereafter it decreases (for H> $H_C$). Thus, it is clear that the positive MR at 2 K arises due to the field induced, first order metamagnetic transition. In the antiferromagnetic region, in presence of the field, the magnetic sublattice antiparallel to the field direction tries to align in the field direction, which leads to an increase in the spin disorder scattering. This contribution to the resistivity increases up to the critical field. These variations give rise to the initial positive MR, whose magnitude decreases at higher fields due to the suppression of the resistivity by the higher field. A comparison of the experimental and theoretical effective moments of HoRhGe gives an indication about the non-zero, but non-localized moment on Rh. This 4d moment being critically dependent on the applied field, the role of additional scattering arising from the spin fluctuations (of Rh moments) on the positive MR cannot be ruled out. Positive MR induced by a metamagnetic transition has not been seen in many systems. A well-known material with somewhat similar behavior is $Tb_5Si_3$ [11]. It has been reported that the MR increases below the metamagnetic transition and decreases above that in $Tb_5Si_3$. It is worth noting from figure 4 that the MR versus T and MR versus H plots are consistent with each other. The positive MR at low temperatures is not seen in MR versus T plot [see figure 4(a)] because the field applied here is higher than the $H_C$ value. The maximum MR in HoRhGe is comparable to some best known rare earth intermetallics with field induced metamagnetic transition such as $SmMn_2Ge_2$ (8%) [12], $Ce(FeRu)_2$ (20%) [13], $Gd_5(Si_2Ge_2)$ (26%) [14], $Gd_5(Si_{1.8}Ge_{2.2})$ (20%) [15].

The magnetocaloric effect (MCE) is an intrinsic property of magnetic materials and manifests as an isothermal magnetic entropy change ($\Delta S_M$) and/or adiabatic temperature change ($\Delta T_{ad}$). In the present case, we have calculated the MCE from both magnetization and heat capacity data. The $\Delta S_M$ from magnetization data [shown in figure 5(a)] has been estimated from the Maxwell's relation

$$\Delta S_M = \int_0^H \left[ \frac{\partial M}{\partial T} \right]_H dH . \qquad (2)$$

The MCE in term of $\Delta S_M$ and $\Delta T_{ad}$ from heat capacity data has been determined using equations

$$\Delta S_M(T,H) = \int_0^T \frac{C(T',H) - C(T',0)}{T'} dT' \qquad (3)$$

$$\Delta T_{ad}(T)_{\Delta H} \cong [T(S)_{H_f} - T(S)_{H_i}]_S \qquad (4)$$

Temperature dependence of $\Delta S_M$ and $\Delta T_{ad}$ calculated from heat capacity data is shown in figure 5(b). The value of the MCE calculated from heat capacity data is found to be 11.1 J/kg K for $\Delta H$ of 50 kOe. The maximum value of $\Delta T_{ad}$ is 4.3 K for 50 kOe. It can be noted that the value of MCE calculated from the magnetization data is slightly lower than the one obtained from the heat capacity data. The refrigerant capacity (RC) defined as in Ref. 16, is found to be 203 J/kg. The observed value of MCE at 50 kOe is comparable to some potential refrigerant materials such as $HoMnO_3$ (13.1 J/kg K at 70 kOe) [17], DySb (15.8 J/kg K) [18], $ErNi_2B_2C$ (9.8 J/kg K) [19], belonging to the same temperature regime.

It has been recognized theoretically and experimentally [20,21] that there is a relationship between $\Delta S_M^{max}$ and H. The magnetic materials with second order transition generally obeys $\Delta S_M^{max} = -kM_s(0)h^{2/3} - S(0,0)$, where h is the reduced field [$h = (\mu_0\mu_B H)/(k_B T_N)$], k is a constant, $M_s(0)$ is the saturation magnetization at low temperatures and $S(0,0)$ is the reference parameter, which may not be equal to zero [20]. We have analyzed $\Delta S_M^{max}$ as a function of $h^{2/3}$ [see inset of figure 5(a)] and found that the sign of $S(0,0)$ is negative, which is expected for second order magnetic transition (SOMT) [20]. The linear dependence of $\Delta S_M^{max}$ versus $h^{2/3}$ implies the strong localization of moments and the second order transition at higher fields in HoRhGe [22]. The fact that the $\Delta S_M^{max}$ is estimated at $T_N$ and in fields larger than the critical field required for the metamagnetic transition justifies the conclusion about the second order transition. It is of importance to note that there is small deviation from the linearity [inset of figure 5(a)] at the lowest field, which is attributed to the first order nature of the transition in low fields. Therefore it is quite evident that the change in the nature of the magnetic transition between low fields and high fields as seen from the magnetization data is also seen from the MCE fit shown in the inset of figure 5(a).

## Conclusions.

In summary, we find that HoRhGe is antiferromagnetic below 5.5 K and shows first order field-induced metamagnetic transition at low fields and below $T_N$. At higher fields, the transitions changes to second order in nature. The negative MR with a magnitude of -25% for 50 kOe was observed near $T_N$. The compound shows positive magnetoresistance at 2 K, which appears as a result of the first order metamagnetic transition. The magnetocaloric effect estimated from M-H-T and C-H-T data was found to be comparable to some potential magnetic refrigerant working in low temperature regime. Field dependence of magnetic entropy change also reflects the change in the nature of magnetic transition as the field in increased.

## Acknowledgments

S.B.G. thanks C.S.I.R., Government of India for fellowship. Authors acknowledge the help rendered by D. Buddhikot in carrying out the resistivity measurement.

**Figure Caption**

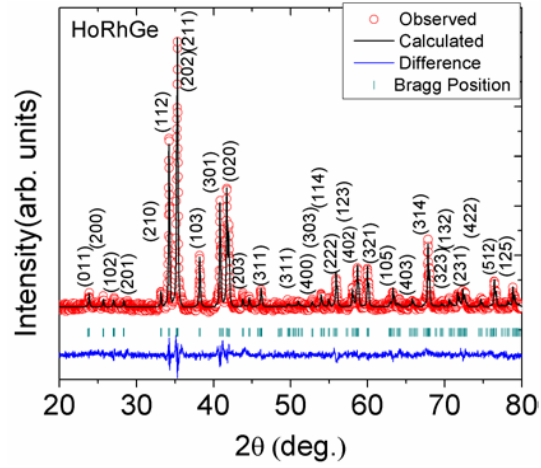

Fig. 1: Powder XRD pattern along with Rietveld refinement for HoRhGe. The plot at the bottom shows the difference between the observed and the calculated patterns.

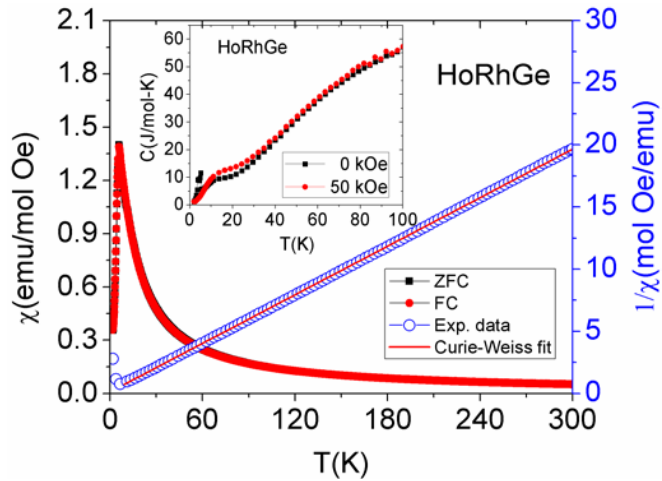

Fig. 2: Temperature dependence of dc magnetic susceptibility (left-hand scale) and the inverse magnetic susceptibility along with the Curie-Weiss fit (right-hand scale) for HoRhGe. The inset shows the heat capacity data with and without field.

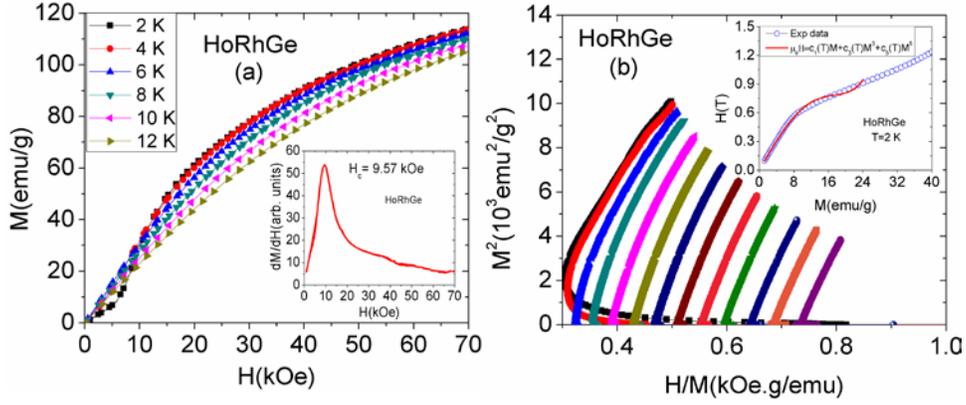

Fig. 3: (a) Magnetization isotherms obtained at selected temperatures in HoRhGe, (b) the Arrott plots at different temperatures. Insets: (a) derivative of magnetization at 2 K, (b) fit of equation of state to the magnetization data.

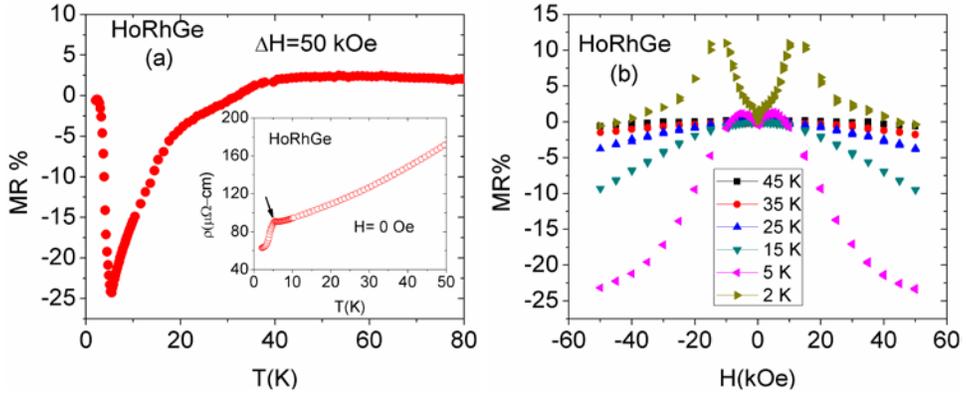

Fig. 4: (a) Temperature dependence of MR in HoRhGe for a field change of 50 kOe, (b) field dependence of MR at different temperatures. The inset in (a) shows the resistivity in an expanded scale for zero field.

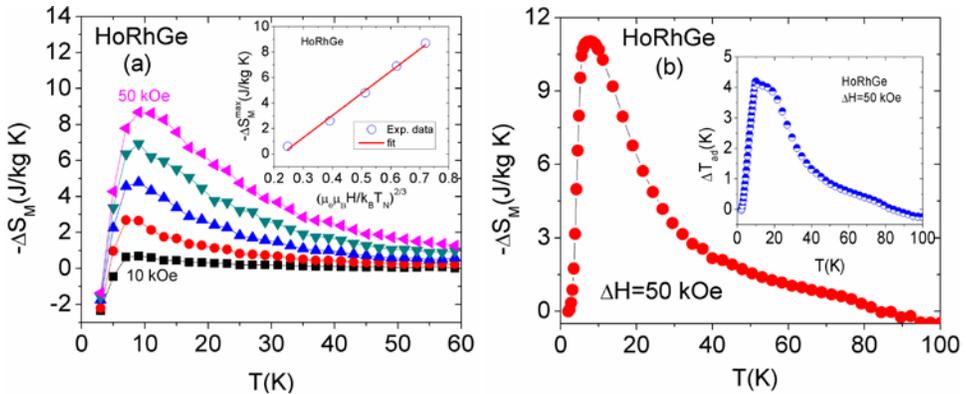

Fig. 5: Temperature dependence of magnetic entropy change in HoRhGe (a) at different fields calculated from the M-H-T data (b) obtained from the C-H-T data at 50 kOe. Insets: (a) field dependence of MCE, (b) temperature dependence of $\Delta T_{ad}$.